\documentclass[twocolumn,11pt]{article}
\usepackage[english]{babel}
\usepackage{a4}
\usepackage{psfig}
\usepackage{times}

\pssilent

\title{Vocal Access to a Newspaper Archive: \\ Design Issues and
Preliminary Investigation}

\author{Fabio Crestani \\ Department of Computing Science \\
University of Glasgow, Scotland \\ email: {\tt fabio@dcs.gla.ac.uk}}

\begin{document}

\maketitle

\begin{abstract}
 
  This paper presents the design and the current prototype
  implementation of an interactive vocal Information Retrieval system
  that can be used to access articles of a large newspaper archive
  using a telephone.  The results of preliminary investigation into
  the feasibility of such a system are also presented.

\end{abstract}

\section{Introduction}
\label{intro}
 
For the last 50 years Information Retrieval has been concerned with
enabling users to retrieve textual documents (most often only
references) in response to textual queries.  With the availability of
faster computers and cheaper electronic storage it is now possible to
store and search the full text of millions of large textual documents
online.  It has been observed that the widespread access to Internet
and the high bandwidth often available to users makes them think that
there is nothing easier than connect and search a large information
repository.  Nevertheless, there are cases in which users may need to
search an information repository without having access to a computer
or to an Internet connection. There are cases in which the only thing
available is a low bandwidth telephone line. We can make a number of
examples of such situations: an automatic booking service for an
airline company that give direct access to recognised customers, a
press archive for journalists on mission abroad (in particular in
developing countries), an automatic telephone based customer support
system, and so on. In all these cases it is necessary to have access
and interact with a system capable not only of understanding the user
spoken query, finding documents and presenting them as speech, but
also capable of interacting with the user in order to better clarify
and specify his information need whenever this is not clear enough to
proceed effectively with the searching. This paper is concerned with
the design issues and the architecture of a prototype of one of such
systems.
 
The paper is organised as follows. Section~\ref{IR+speech} introduces
the difficult marriage between IR and speech. Section~\ref{SIRE} gives
the background of this work and explains its final objective: the
Interactive Vocal Information Retrieval System. Section~\ref{proto}
reports on the current state of the implementation of the prototype
system, while section~\ref{eval} briefly describes some interesting
findings of the feasibility study.

\section{Information Retrieval and Speech}
\label{IR+speech}

{\em Information Retrieval\/} (IR) is the branch of computing science
that aims at storing and allowing fast access to a large amount of
multimedia information, like for example text, images, speech, and so
on~\cite{vanRijsbergen79}.  An {\em Information Retrieval System\/} is
a computing tool that enables a user to access information by its
semantic content using advanced statistical, probabilistic, and/or
linguistic techniques.

Most current IR systems enable fast and effective retrieval of textual
information or documents, in collections of very large size, sometimes
containing millions of documents. The retrieval of multimedia
information, on the other hand, is still an open problem.  Very few IR
systems capable of retrieving multimedia information by its semantic
content have been developed. Often multimedia information is retrieved
by means of an attached textual description.

The marriage between IR and speech is a very recent event. IR has been
concerned for the last 50 years with textual documents and queries. It
is only recently that talking about multimedia IR has become possible.
Progress in speech recognition and synthesis~\cite{Holmes88} and the
availability of cheap storage and processing power have made possible
what only a few years ago was unthinkable.
 
The association between IR and speech has different possibilities:
 
\begin{itemize}
  
  \item textual queries and spoken documents;
  
  \item spoken queries and textual documents;
 
  \item spoken queries and spoken documents.
 
\end{itemize}
 
In the large majority of current IR systems capable of dealing with
speech, the spoken documents or queries are first transformed into
their textual transcripts and then dealt by the IR system with
techniques that are derived from those used in normal textual IR.
 
The retrieval of spoken documents using a textual query is a fast
emerging area of research (see for example~\cite{Sparck-Jones&96}). It
involves an efficient, more than effective, combination of the most
advanced techniques used in speech recognition and IR. The increasing
interest in this area of research is confirmed by the inclusion, for
the first time, of a retrieval of spoken documents retrieval track in
the TREC-6 conference~\cite{Voorhees&97}. The problem here is to
devise IR models that can cope with the large number of errors
inevitably found in the transcripts of the spoken documents. Models
designed for retrieval of OCRed documents have proved useful in this
context~\cite{Mittendorf&96}.  Another problem is related to the fact
that, although IR models can easily cope with the fast searching of
large document collections, fast speech recognition of a large number
of long spoken documents is a much more difficult task. Because of
this, spoken documents are converted into textual transcripts
off-line, and only the transcripts are dealt by the IR system.
 
The problem of retrieving textual documents using a spoken query may
seem easier than the previous one, because of the smaller size of the
speech recognition task involved.  However, it is not so.  While the
incorrect or uncertain recognition of an instance of a word in a long
spoken document can be compensated by its correct recognition in some
other instances, the incorrect recognition of a word in a spoken query
can have disastrous consequences. Queries are generally very
short\footnote{There is an on-going debate about realistic query
lengths. While TREC queries are on average about 40 words long, Web
queries are only 2 words long on average. This recently motivated the
creation in TREC of a ``short query'' track, to experiment with
queries of more realistic length.} and the failing of recognising a
query word, or worse, the incorrect recognition of a query word will
fail to retrieve a large number of relevant documents or wrongly
retrieve a large number of non-relevant documents. 
 
The retrieval of spoken queries in response to spoken documents is a
very complex task and is more in the realm of speech processing than
IR, although IR techniques could be useful. Speech recognition and
processing techniques can be used to compare spoken words and
sentences in their raw form, without the need of generating textual
transcripts. We will not address this issue here.

\section{The Interactive Vocal Information Retrieval System}
\label{SIRE}

\begin{figure}
        \centerline{\psfig{figure=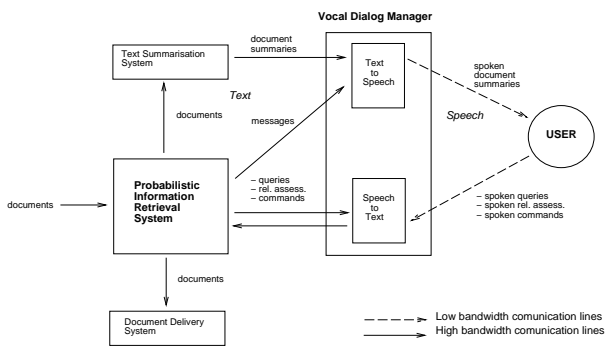,width=8cm}}
        \caption{Schematic view of the IVIRS prototype}
        \label{fig:sys-spec}
\end{figure}

In the second half of 1997, at Glasgow University, we started a project
on the sonification of an IR environment. The project is funded by the
European Union under the Training and Mobility of Researchers (TMR)
scheme of the Fourth Framework. The main objective of the project is
to enable a user to interact (e.g.~submit queries, commands, relevance
assessments, and receive summaries of retrieved documents) with a
probabilistic IR system over a low bandwidth communication system,
like for example a telephone line. An outline of the system
specification is reported in the figure \ref{fig:sys-spec}.
 
The {\em interactive vocal information retrieval system\/} (IVIRS),
resulting from the ``sonification'' of a probabilistic IR system, has
the following components:
 
\begin{itemize}
 
  \item a {\em vocal dialog manager\/} (VDM) that provides an
  ``intelligent'' speech interface between user and IR system;
  
  \item a {\em probabilistic IR system\/} (PIRS) that deals with the
  probabilistic ranking and retrieval of documents in a large textual
  information repository;
 
  \item a {\em document summarisation system\/} (DSS) that produces a
  summary of the content of retrieved documents in such a way that the
  user will be able to assess their relevance to his information need;
 
  \item a {\em document delivery system\/} (DDS) that delivers
  documents on request by the user via electronic mail, ftp, fax, or
  postal service.
 
\end{itemize}

It is important to emphasise that such a system cannot be developed
simply with off the shelf components. In fact, although some
components (DSS, DDS, and the Text-to-Speech module of the VDM) have
already been developed in other application contexts, it is necessary
to modify and integrate them for the IR task.

The IVIRS prototype works in the following way. A user connects to the
system using a telephone. After the system has recognised the user by
means of a username and a password (to avoid problems in this phase we
devised a login procedure based on keying in an identification number
using a touch tone), the user submit a spoken query to the system. The
VDM interact with the user to identify the exact part of spoken
dialogue that constitutes the query. The query is then translated into
text and fed to the PIRS. Additional information regarding the
confidence of the speech recognisers is also fed to the PIRS. This
information is necessary in order to limit the effects of wrongly
recognised words in the query, Additionally, an effective interaction
between the system and the user can also help to solve this problem.
The system could ask the user for confirmation in case of a uncertain
recognition of a word, asking to re-utter the word or to select one of
the possible recognised alternatives.  The PIRS searches the textual
archive and produces a ranked list of documents, and a threshold is
used to find the a set of document regarded as surely relevant (this
feature can be set in the most appropriate way by the user). The user
is informed on the number of documents found to be relevant and can
submit a new query or ask to inspect the documents found. Documents in
the ranked list are passed to the DSS that produces a short
representation of each document that is read to the user over the
telephone by the Text-to-Speech module the VDM. The user can wait
until a new document is read, ask to skip the document, mark it as
relevant or stop the process all together. Marked documents are stored
in retrieved set and the use can proceed with a new query if he wishes
so. A document marked as relevant can also be used to refine the
initial query and find additional relevant documents by feeding it
back to the PIRS. This relevance feedback process is also useful in
case of wrongly recognised query words, since the confidence values of
query words increase if they are found in relevant documents. This
interactive process can go on until the user is satisfied with the
retrieved set of documents. Finally, the user can ask the documents in
the retrieved set to be read in their entirety or sent to a known
addressed by the DDS.

\subsection{The Vocal Dialog Manager}

The human-computer interaction performed by the {\em VDM\/} is not a
simple process that can be done by off the shelf devices. The VDM
needs to interact in an ``intelligent'' way with the user and with the
PIRS in order to understand completely and execute correctly the
commands given by the user. In fact, while the technology available to
develop the speech synthesis (Text-to-Speech) module is sufficient for
this project (but see section~\ref{spoken-rel}), the technology
available for the speech recognition (Speech-to-Text) module is
definitely not. On one hand, the translation of speech into text is a
much harder task than the translation of text to speech, in particular
in the case of continuous speech, speaker independent, large
vocabulary, noisy channel speech recognition. On the other hand, it is
necessary to take into consideration the three possible forms of
uncertainty that will be present in the IR environment by adding a
vocal/speech component:
 
\begin{enumerate}
  
  \item the uncertainty related to the speech recognition process;
  
  \item the uncertainty given by the word sense ambiguity present in
  the natural language;
  
  \item the uncertainty related to the use of the spoken query in the
  retrieval process.
 
\end{enumerate}
 
In order to deal with these different forms of uncertainty we need not
only to developed a advanced retrieval model for the PIRS, but also
develop a model of the user-VMD-PIRS interaction that encompasses a
language model, an interaction model and a translation model. This
later part of the project is still at the early stages of development
and will not be presented here. In this context we also make use of
the results of previous work in this area, although in rather
different applications (see for example~\cite{Peckham91,Bernsen&97}.

Currently we are building a model of the telephone interactions
between users and PIRS analysing the results of a study carried out
using the ``Wizard of Oz'' technique. The Wizard of Oz technique is a
way of simulating human-computer interfaces. Unknown to the user, a
human ``wizard'' performs some of the functions of the computer such
as responding to the user's spoken input. The technique is commonly
used where the technology for running the interface does not yet exist
or is not yet sophisticated enough to be used in real time. Examples
include experiments where the recognition vocabulary is within current
capabilities (e.g.~sequences of digits) but the recognition
performance required is beyond current capabilities.

The particular Wizard of Oz simulation we are currently using for the
design of the VDM incorporates a statistical model of word
recognition errors, whereby a realistic distribution of speech recogniser
errors can be generated at any desired overall accuracy level.

A limitation to the realism of this form of simulation is that
recognition performance depends only on the content of the user's
input, and not on its quality (clarity of speaking, background noise
etc) as it would with a real recogniser. This is particularly relevant
in the case of word-spotting, where the recogniser is designed to pick
out instances of keywords embedded in arbitrary speech. Typically the
accuracy of a real word-spotter is better for isolated keyword
utterances than for embedded ones. To address this, a
second-generation simulation method is currently being developed, in
which the wizard's input (giving the keyword content of the utterance)
is combined with acoustic information extracted automatically from the
speech signal. This will enable us to design appropriate error
recovery strategies whenever one or more words in the spoken query are
below a certain threshold of recognition.

The VMD has two sub-components: a Speech-to-Text module and a
Text-to-Speech module.

The {\em Speech-to-Text module\/} is arguably the most important and
the most problematic module of the VMD. Speech recognition has been
progressing very rapidly in the last few years~\cite{Keller94} and
results are improving day by day, in particular for speaker dependent
systems. There is already a number of system commercially available
that guarantee quite impressive performance once they have been
properly trained by the user (e.g.~Dragon Naturally Speaking, IBM Viva
Voce, etc.). The user also needs to teach how to recognise words that
are not part of the recogniser's vocabulary. The situation is not so
good with speaker independent continuous speech recognition systems
over the telephone, although a number of studies have achieved some
acceptable results~\cite{Kao&94,Peckham91}. In this context we do not
intend to develop any new speech recognition system. This is a
difficult area of research for which we do not have necessary
know-how. Instead, we make use of available state-of-the-art
technology and try to integrate it in our system. The aim is to enable
the PIRS to deal with the uncertainty related to the spoken query
recognition process, integrating this uncertainty with the classical
uncertainty of intrinsic in the IR process.  Two strategies are
currently being experimented with:

\begin{itemize}
  
  \item use the speech recogniser's ``confidence'' in recognising a
  word;
  
  \item merging the output of a number of speech recognisers to
  compensate for errors of single speech recognisers and create an
  ``combined confidence'' in recognising a word.

\end{itemize}

In section~\ref{spoken-queries} we report some initial results on the
use of these techniques. 

The {\em Text-to-Speech module\/} uses the state-of-the-art technology
in speech synthesis~\cite{Keller94}. We carried out a survey and an
initial testing of available speech synthesis systems. In
section~\ref{spoken-rel} we report some initial results of an
experimentation into the user's perception of relevance of spoken
documents. In the experiments carried out we used to different
commercial speech synthesis systems.

\subsection{The Probabilistic Information Retrieval System}

\begin{figure}
        \centerline{\psfig{figure=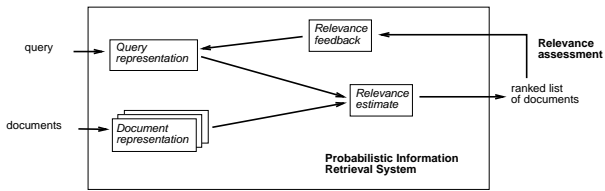,width=8cm}}
        \caption{Schematic view of a probabilistic IR system.}
        \label{fig:irs}
\end{figure}

The {\em PIRS\/} performs a ranking of the document in the collection
in decreasing order of their estimated probability of relevance to the
user's query. Past and present research has made  use of formal
probability theory and statistics in order to solve the problems of
estimation~\cite{Crestani&98c}. An schematic example of how a
probabilistic IR system works is depicted in figure~\ref{fig:irs}.
 
Currently there is no PIRS with speech input/output capabilities. The
adding of these capabilities to a PIRS is not simply a matter of
fitting together a few components, since there needs to be some form
of feedback between PIRS and VMD so that the speech input received
from the user and translated into text can be effectively recognised
by the PIRS and used. If, for example, a word in a spoken query is not
recognised by the PIRS, but is very close (phonetically) to some other
word the PIRS knows about and that is used in a similar context
(information that the PIRS can get from statistical analysis of the
word occurrence), then the PIRS could suggest to the user to use this
new word in the query instead on the unknown one.

Additionally, it is generally recognised in IR that improvement in
effectiveness can be achieved by enhancing the interaction between
system and user.  One of the most advanced techniques used for this
purpose is {\em relevance feedback\/} (RF). RF is a technique that
allows a user to interactively express his information requirement by
adapting his original query formulation with further information
\cite{Harman92a}.  This additional information is often provided by
indicating some relevant documents among the documents retrieved by
the system. When a document is marked as relevant the RF process
analyses the text of the document, picking out words that are
statistically significant, and modifies the original query
accordingly. RF is a good technique for specifying an information
need, because it releases the user from the burden of having to think
up of words for the query and because it limits the effects of errors
in the recognition of query words.  Instead the user deals with the
ideas and concepts contained in the documents. It also fits in well
with the known human trait of ``I don't know what I want, but I'll
know it when I see it''.  RF can also be used to detect words that
were wrongly recognised by the VMD.  In fact if, for example, a query
word uttered by the user never appears in documents that the user
points out to be relevant, while another word similarly spelled (or
pronounced) often occurs, then it is quite likely (and we can evaluate
this probability) that the VMD was wrong in recognising that word and
that the word really uttered by the user is the other word.  The
overall performance of the interactive system can be enhanced also
using other similar forms of interactions between PIRS and VMD that
are current object of study.

\subsection{The Document Summarisation System}

The {\em DSS\/} performs a query oriented document summarisation aimed
at stressing the relation between the document content and the query.
This enables the user to assess in a better and faster way the
relevance of the document to his information need and provide correct
feedback to the PIRS on the relevance of the document.  Query oriented
document summarisation attempts to concentrate users' attention on the
parts of the text that possess a high density of relevant information.
This emerging area of research has its origins in methods known as
passage retrieval (see for example~\cite{Callan94}). These methods
identify and present to the user individual text passages that are
more focussed towards particular information needs than the full
document texts. The main advantage of these approaches is that they
provide an intuitive overview of the distribution of the relevant
pieces of information within the documents. As a result, it may be
easier for users to decide on the relevance of the retrieved documents
to their queries.  
 
The summarisation system employed in IVIRS is the one developed by
Tombros and Sanderson at Glasgow University~\cite{Tombros&98}.  The
system is based on a number of sentence extraction methods that
utilise information both from the documents of the collection and from
the queries used. A thorough description of the system can be found
in~\cite{Tombros&98} and will not be reported here. In summary, a
document passes through a parsing system, and as a result a score for
each sentence of the document is computed. This score represents the
sentence's importance for inclusion in the document's summary. Scores
are assigned to sentences by examining the structural organisation of
each document (to distinguish word in the title or in other important
parts of the document, for example), and by utilising within-document
word frequency information. The document summary is then generated by
selecting the top-scoring sentences, and outputting them in the order
in which they appear in the full document. The summary length is a
fraction of the full document's length. For the documents used in our
experimentation the summary length was set at 15\% of the document's
length, up to a maximum of five sentences.

\subsection{The Document Delivery System}

The {\em DDS\/} performs the delivery of all or parts of the
document(s) requested by the user. The user can decide the way and
format of the document delivery; this information is usually stored in
a user profile, so that the user does not need to give this
information every time he uses the system. Documents can be delivered
by voice (read in their entirety to the user through the telephone),
electronic mail, postal service, or fax; if delivered by electronic
mail a number of different document formats are available, like for
example, PDF, postscript, RDF, or ASCII.

\section{Prototype Implementation}
\label{proto}

The implementation of the prototype system outlined in the previous
sections requires, as a first step, a careful choice of some already
existing software components: a speech recognition system, a speech
synthesis system, a probabilistic IR system, and a document
summarisation system. This calls for a survey of the state-of-the-art
of several different areas of research some of which are familiar to
us, while others are new to us. A second step involves the development
of a model for the VDM and of its interaction with the other
components. The prototype implementation of the overall system
requires a careful tuning and testing with different users and
in several different conditions (noisy environment, foreign speaker,
etc.).

The prototype implementation of IVIRS is still in progress. A ``divide
et impera'' approach is currently being followed, consisting of
dividing the implementation and experimentation of IVIRS in the
parallel implementation and experimentation of its different
components. The integration of the various components will be the last
stage.  Currently we have implemented and experimented with the DSS,
the Text-to-Speech and Speech-to-Text modules of the VDM, and the DDS.
We are currently developing the PIRS, and the VDM.

\section{Initial Evaluation Results}
\label{eval}

In this section we report on the initial results found experimenting
with the DSS, and the Text-to-Speech and Speech-to-Text modules of the
VDM. This constitutes part of a feasibility study of the system. In the
pilot implementation of the IVIRS prototype we use a collection of
newspaper articles, in particular the TREC Wall Street Journal
collection.  

\subsection{Experimentation of the Document Summarisation System and
Text-to-Speech Module}
\label{spoken-rel}

One of the underlying assumptions of the design and development of the
IVIRS system is that a user would be able to assess the relevance of a
retrieved document by hearing a synthesised voice reading a brief
description of its semantic content through a telephone line. This is
essential for an effective use of the system. In order to test the
validity of this assumption we carried out a series of experiments
with the DSS and Text-to-Speech module of the IVIRS. The aim of this
experimentation was to investigate the effect that different forms of
presentation of document descriptions have on users' perception of the
relevance of a document.  In a previous study, Tombros and
Sanderson~\cite{Tombros&98} used document titles, and automatically
generated, query biased summaries as document descriptions, and
measured user performance in relevance judgements when the
descriptions were displayed on a computer screen and read by the
users. The results from that study were used in this experiment, and
compared to results obtained when users are listening to the document
descriptions instead of reading them. Three different ways of auditory
transmission are employed in our study: document descriptions are read by
a human to the subjects, read by a human to the subjects over the
telephone, and finally read by a speech synthesis system over the
telephone to the subjects.  The objective was that by manipulating the
level of the independent variable of the experiment (the form of
presentation), we could examine the value of the dependent variable of
the experiment (the user performance in relevance judgements). We also
tried to prove that any variation in user performance between the
experimental conditions was to be attributed only to the change of
level of the independent variable. In order to be able to make such a
claim, we had to ensure that the so-called ``situational variables''
(e.g.~background noise, equipment used, experimenter's behaviour) were
held constant throughout the experimental procedure. Such variables
could introduce bias in the results if they systematically changed
from one experimental condition to another~\cite{Miller84}.
 
In order to be able to use the experimental results reported
in~\cite{Tombros&98}, the same task was introduced in our design:
users were presented with a list documents retrieved in response to a
query, and had to identify as many documents relevant to that
particular query as possible within 5 minutes. The information that
was presented for each document was its title, and its automatically
generated, query oriented description. Moreover, we used exactly the same
set of queries (50 randomly chosen TREC queries), set of retrieved
documents for each query (the 50 top-ranked documents), and document
descriptions as in~\cite{Tombros&98}. Queries were randomly allocated to
subjects by means of a draw, but since each subject was presented with
a total of 15 queries (5 queries for each condition) we ensured that
no query was assigned to a specific user more than once.  A group
consisting of 10 users was employed in the experimental procedure. The
population was drawn from postgraduate students in computer science.
All users performed the same retrieval task described in the previous
paragraph under the three different experimental conditions.

The experiment involved the presentation of document descriptions to
subjects in three different forms, all of which were of an auditory
nature. In two of the experimental conditions the same human read the
document descriptions to each subject, either by being physically in
the same room (though not directly facing the subject), or by being
located in a different room and reading the descriptions over the
telephone. In the last experimental condition a speech synthesiser
read the document description to the user over the telephone.  User
interaction with the system was defined in the following way: the
system would start reading the description of the top ranked document.
At any point in time the user could stop the system and instruct it to
move to the next document, or instruct it to repeat the current
document description. If none of the above occurred, the system would
go through the current document description, and upon reaching its end
it would proceed to the next description.
 
\footnotesize
\begin{table}[ht]
\begin{center}
\begin{tabular}{|l|c|c|c|c|} \hline \hline
       & S  & V & T & C \\ \hline
Avg. P. \% & 47.15  & 41.33 &  43.94 & 42.27 \\
Avg. R. \% &  64.84  & 60.31 &  52.61 & 49.62 \\
Avg. T. (sec.) & 17.64 & 21.55 &  21.69 & 25.48 \\
\hline \hline
\end{tabular}
\caption{Average precision, recall, and time in the four assessment
conditions.}
\label{tab:averages}
\end{center}
\end{table}
\normalsize

Table~\ref{tab:averages} reports the results of the user relevance
assessment in terms of precision, recall and average time for all four
conditions: on screen description (S), read description (V),
description read over the telephone (T), and computer synthesised
description read over the telephone (C).  These results show how users
in condition S performed better in terms of precision and recall and
were also faster. Recall and average time slowly decreased from S to
C, although some of these differences are not statistically
significant (i.e.~the average time of V and T). We were surprised to
notice that precision was slightly higher for condition T than for
conditions V or C.  Users tended to be more concentrated when hearing
the descriptions read over the telephone than by the same person in
the same room, in front of them, but this concentration was not enough
when the quality of the voice was getting worse.  Nevertheless, the
difference in precision between conditions S and C was not so large
(only about 5\%) to create unsolvable problems for a telephone based
IR system.

A difference that was certainly significant was in the average time
taken to assess the relevance for one document. The difference between
the condition S and C was quite large and enabled a user to assess on
average, in the same amount of time (5 minutes), 70\% more documents
in condition S than in condition C (22 documents instead of 13). A
sensible user would have to evaluate if it is more cost effective, in
terms of time connected to the service, to access the system using
computer and modem and looking at the documents on the screen, than
accessing the system using a telephone.  Nonetheless, difference in
the average time taken to assess the relevance for one document were
very subjective, in fact, we could notice that some users were slow
whatever the condition, while other were always fast.

These preliminary results enable us to conclude that a IVIRS system is
indeed feasible and, provided we solve the issued related to the
correct recognition of the user query, we can expect to develop a
system that could be useful.

\subsection{Experimentation of the Speech-to-Text Module}
\label{spoken-queries}

In~\cite{Crestani&97b,Sanderson&98} we presented the results of the
experimentation of a number of techniques for the retrieval of spoken
documents. In particular, two techniques were experimented, both
attempting to use confidence values generated in the speech
recognition process to deal with the uncertainty related to the spoken
words in the documents. In one set of experiments we used confidence
values generated by the speech recogniser, in another set of
experiments we generated this confidence values by merging the
transcripts of a number of different recognisers. While the first
strategy was not successful due to our ignorance of the way the
confidence values are generated by a speech recogniser, using the
simple strategy of merging the transcripts of spoken documents of
different recognisers showed most promise.

In the context of the SIRE project we attempted to use the same
technique of merging different transcripts with spoken queries. The
challenge here is due to the fact the queries are usually much shorter
than documents and errors in the recognition of words in queries cause
more damage to retrieval effectiveness than errors in the recognition
of words in documents.

Although this experimentation is still going on, the initial results
were not encouraging. We experimented with merging the transcripts of
spoken queries originated by two and three different speech
recognition systems. The improvement in word recognition and
confidence obtained over the use of a single speech recognition
system, although considerable, was not enough to increase the level of
effectiveness of a PIRS. The major problem was caused by the inclusion
in the transcript of queries of words wrongly recognised. Although the
event of words being wrongly recognised by more that a recogniser was
rare, resulting in wrongly recognised words having usually low
confidence levels, the inclusion of any such word in the query
transcript had disastrous effects on the effectiveness of the
retrieval process. The ranking produced by the IR system was highly
influenced by the wrongly recognised words since this were quite often
rare words and therefore had high indexing weights.  Instead, the
effect of missing a word from the query transcript was not
detrimental, but this result may not be entirely fair as very few of
the queries we used had words outside recognisers' vocabulary.  More
``realistic'' queries containing many proper nouns might produce
different results and require an alternative approach: for example, a
recogniser using both word and sub-word unit recognition. An
alternative strategy could consist in using a PIRS employing a query
expansion technique to expand, from a text corpus, unrecognised query
words with those in the recognisers' vocabulary (using, for example,
Local Context Analysis~\cite{Xu&96}). We are currently experimenting
with the latter technique.

\section{Conclusions}

In this paper we outlined the design of an interactive vocal IR
system. We also reported on some initial experimentation which
highlights the complexity of the implementation of such a system. The
work presented here is still in progress and the full implementation
of the system is under way. We expect to be able to have a working
prototype system very soon. The evaluation of such system and, in
particular, of the user interaction will constitute an exciting area
of research.

\section*{Acknowledgements}

The author is currently supported by a ``Marie Curie'' Research
Fellowship from the European Union. 

The experimentation reported in section~\ref{spoken-rel} was carried
out with the help of Anastasios Tombros.

\bibliography{/users/ir/fabio/Bib/general,/users/ir/fabio/Bib/unread} 

\bibliographystyle{plain}
 
\end{document}